\documentclass[%
 reprint,
 amsmath,amssymb,
 aps,
]{revtex4-2}

\usepackage{graphicx}
\usepackage{dcolumn}
\usepackage{bm}
\renewcommand{\theequation}{\arabic{equation}}

\begin{document}
\title{Limits of funneling efficiency in non-uniformly strained 2D semiconductors}

\author{Moshe G. Harats}
\affiliation{Racah Institute of Physics, The Hebrew University of Jerusalem,  Jerusalem 9190401, Israel}

\email{moshe.harats@mail.huji.ac.il}
\author{Kirill I. Bolotin}%
 
\affiliation{Department of Physics, Freie Universit\"{a}t Berlin, 14195 Berlin, Germany}

\begin{abstract}
Photoexcited electron-hole pairs (excitons) in transition metal dichalcogenides (TMDC) experience an effective force when these materials are non-uniformly strained. In the case of strain produced by a sharp tip pressing at the center of a suspended TMDC membrane, the excitons are transported to the point of the highest strain at the center of the membrane. This effect, exciton funneling, can be used to increase photoconversion efficiency in TMDC, to explore exciton transport, and to study correlated states of excitons arising at their high densities. Here, we analyze the limits of funneling efficiency in realistic device geometries. The funneling efficiency in realistic monolayer TMDCs is found to be low, $ <5 \;\%$  both at room and low temperatures. This results from dominant diffusion at room temperature and short exciton lifetimes at low temperatures. On the other hand, in TMDC heterostructures with long exciton lifetimes the funneling efficiency reaches $\sim 50\;\%$  at room temperature, as the exciton density reaches thermal equilibrium in the funnel. Finally, we show that Auger recombination limits funneling efficiency for intense illumination sources.

\end{abstract}

\maketitle

\section{\label{sec:intro}Introduction}
Inducing a controlled mechanical strain (strain engineering) is crucial in improving the performance of semiconductor devices. The most common use of strain engineering is to enhance the carrier mobility in Silicon MOSFET transistors \cite{Manasevit1982,People1984a}. Strain is also used for dynamical control of the bandgap of semiconductors by surface acoustic waves \cite{Lazic2014}. With the emergence of Van der Waals (2D) materials that followed the discovery of monolayer graphene \cite{Novoselov2004ElectricFilms}, there is increased interest in the role of strain in 2D materials. The effects of strain are especially strong in semiconducting materials from the group of transition metal dichalcogenides (TMDCs). In monolayer TMDCs, strain changes the nature of the bandgap from direct to indirect \cite{Conley2013}, affects excitonic linewidths \cite{Niehues2018}, and produces pseudomagnetic fields \cite{Wang2017Valley-WSe2}. In addition, TMDCs sustain strain higher than $10\;\%$ \cite{Bertolazzi2011} and possess a relatively high Young's modulus of $\sim 100\;GPa$ \cite{Harats2020DynamicsWS2b}. These features make it interesting to consider strain-engineering of TMDCs towards potential applications in flexible electronic and opto-electronic devices. 

In a pioneering work, Feng et al. proposed one particular application of strain engineering to increase TMDC photocoversion efficiency \cite{Feng2012}. They considered a suspended monolayer TMDC on top of a circular hole. The membrane is indented at the center by a nanometer-sized electrically conductive tip, generating a non-uniform strain profile in it. Under strain, the band-gap is reduced as a result of diminishing overlap of the orbitals between adjacent atoms \cite{Shi2013}. For the case of uniaxial strain, the reduction has been experimentally measured to be $\sim 50\;meV/\%$ for monolayer WS$_2$ \cite{Conley2013,Niehues2018}. The spatially non-uniform strain results in a spatially non-uniform band-gap resembling a "funnel" (see Fig. \ref{fig:fig1_sketch}b). Photogenerated excitons are created throughout the structure under illumination. The excitons generated at the far end of the device experience an effective force transporting them to the deepest point of the potential well funnel at the center of the suspended monolayer. Finally, the excitons accumulated in the region of the maximum strain under the tip are separated into charge carriers and extracted into the tip as electrical current. Feng et al. suggested that this approach allows for efficient collection of excitons throughout the TMDC and can be used to improve the efficiency of TMDC-based solar cells \cite{Feng2012}. In addition, exciton funneling is interesting as an experimental tool to generate large density of excitons and to explore transport of excitons under the influence of strong forces \cite{Harats2020DynamicsWS2b}. This proposal has lead to  works exploring signatures of exciton funneling in various experimental geometries \cite{Castellanos-Gomez2013,Tyurnina2019}.

In a recent study, we have realized for the first time the experimental geometry proposed in Ref. \onlinecite{Feng2012} using a home-built all-electrical AFM setup that can operate at ambient and vacuum conditions \cite{Harats2020DynamicsWS2b}. We then used photoluminescence (PL) spectroscopy to interrogate the dynamics of photo-generated excitons under the non-uniform strain profile generated by the applied force of the AFM tip. Surprisingly, we found that not only "funneling" of excitons is negligible, but that the dominant physical process affecting PL spectra was an efficient conversion of excitons to negatively charged trions. We revealed that the diffusion of the excited states (either excitons or trions) is limiting the "funneling" efficiency. Moreover, at least at room temperature, we found that the PL spectra of our devices are described with good accuracy by a model where excitons and trions are static, i.e. recombine radiatively from the same spatial position where they were excited. This has been further confirmed in a different setup in which the non-uniform strain was induced by pressurizing the TMDCs 
monolayer \cite{Kovalchuk2020NeutralWS2b}.

The question remains, can "funneling" be efficient? Are there device configurations, material parameters, or experimental conditions under which a significant (e.g. $50\;\%$) fraction of photoexcited excitons reaches the strain funnel center?  Here, we analyze in depth the multi-dimensional parameter space of experimental parameters (temperature, material composition, exciton mobility, etc.) and we find experimentally feasible conditions leading to efficient funneling.

\section{Model\label{sec:model}}

\begin{figure}[hbtp]
    \centering
    \includegraphics[width=0.9\columnwidth]{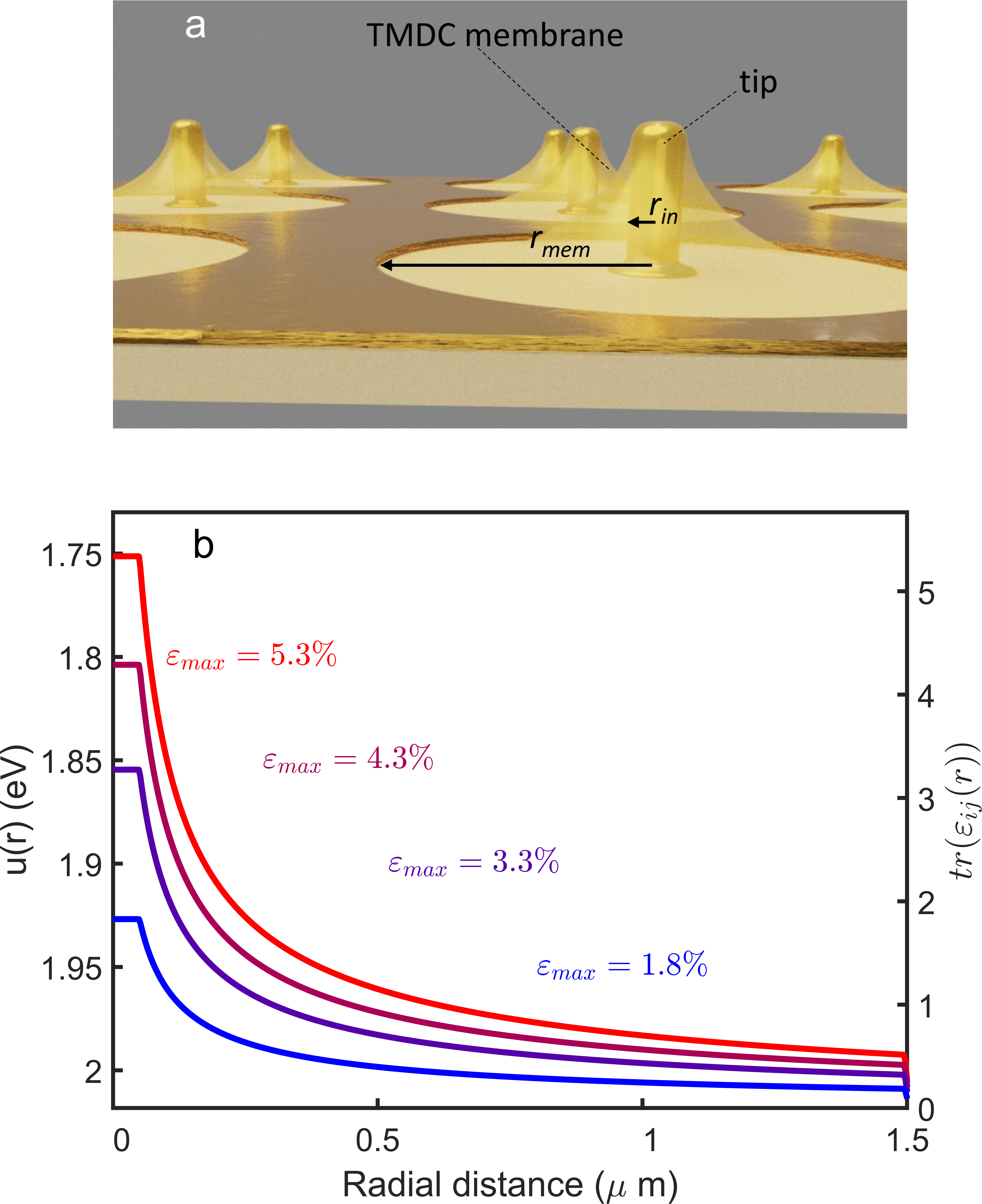}
\caption{(a) A sketch of the funneling device. A TMDC monolayer is deposited onto sharp metallic electrodes inducing non-uniform strain in it. The monolayer is illuminated from the top. Photogenerated excitons are subjected to a drift force which transports them to the region of high strain at the apex of the tip. (b) The local strain, $tr(\varepsilon_{ij})(r)$, and bandgap, $u$, calculated as a function of distance from the membrane's center $r$ for different maximum strain ($\varepsilon_{max}$) values.}
    \label{fig:fig1_sketch}
\end{figure}

We will consider funneling of excitons in the device shown in Fig. \ref{fig:fig1_sketch}a. In that device, a TMDC monolayer is deposited onto a "bed of nails" \cite{Guinea2010}: an array of electrically contacted sharp spherical tips. A circular electrode around each tip serves as a counter-electrode. The strain profile is determined by the height of the tips. Similar devices have already been fabricated using scalable nanofabrication workflows \cite{Jiang2017}. Moreover, the device geometry is similar to the one examined earlier in our experimental work where the strain was induced by an AFM tip \cite{Harats2020DynamicsWS2b}. Therefore, in our model we will use many of the experimental parameters measured in that work.

We will analyze funneling of excitons in the device shown in Fig. \ref{fig:fig1_sketch}a in three steps. The first step is the calculation of the strain tensor as a function of the position. Borrowing the parameters from our experimental work in Ref. \onlinecite{Harats2020DynamicsWS2b}, we will assume that radius of the membrane is $r_{mem}=1.5\;\mu m$ and the radius of the tip is $r_{in}=50\;nm$. From the family of TMDC materials, we will concentrate on monolayer WS$_2$, the material for which excitonic lifetimes and diffusion lengths have been thoroughly measured \cite{Kulig2018,Zipfel2020ExcitonDisorder}. We assume circular symmetry, that the monolayer is clamped at the edges, and that the contact area between the tip and the monolayer is $r_{in}$ for every force exerted by the tip. The strain profiles in our devices are controlled by changing the height of the central tip (see Fig. \ref{fig:fig1_sketch}b). To analyze only  realistic strain distributions, we assumed that the maximum strain $\varepsilon_{max}$ (that is reached at the position of the tip) remains smaller than $6\;\%$. Above this strain, the typical devices in Ref. \onlinecite{Harats2020DynamicsWS2b} ruptured.  We assume a WS$_2$ Young's modulus of $E=100\;GPa$ \cite{Harats2020DynamicsWS2b}, a Poisson ratio of $\nu=0.25$, and that these parameters are independent of applied forces. Using all of the assumptions above, we calculate the strain distribution in WS$_2$ as a function of the parameter $\varepsilon_{max}$ using the method given by Ref. \onlinecite{Vella2017}.

Next, we find the local band-gap $u(r)$ from the spatially dependent strain tensor $\varepsilon_{ij}(r)$. We assume that $u(r)=E_0-0.05\cdot tr(\varepsilon_{ij}(r))$ where $E_0=2.08 \;eV$ is the band-gap energy at zero strain for WS$_2$ and  $tr(\varepsilon_{ij}(r))$ is the trace of the cylindrical strain tensor. The bandgap reduction with strain,  $\sim 50\;meV/\%$ is estimated from available literature \cite{Conley2013,Niehues2018}.

Under these assumptions, the spatially dependent band-gap becomes:
\begin{equation}\label{eq:bandgap}
    u(r)=\left\{
        \begin{array}{c c}
           E_0-0.05\cdot\varepsilon_{max} & r\leq r_{tip}\\
        E_0-0.05\cdot(\varepsilon_{rr}(r)+\varepsilon_{\theta\theta}(r)) & r_{tip}\leq r \leq r_{mem}\\
        E_0 & r_{mem} \leq r
        \end{array}\right.
\end{equation}
Here $\varepsilon_{rr}(r)$ and $\varepsilon_{\theta\theta}(r)$ are the spatially-dependent radial and angular strain tensor components, respectively. Due to the vanishing thickness of the 2D material we neglect the bending rigidity of the monolayer and hence $\varepsilon_{zz}=0$ \cite{Vella2017}. 

In the final step, we use the spatially dependent band-gap calculated using Eq. \ref{eq:bandgap} to determine the equilibrium exciton density $n(r)$. We solve the following axis-symmetric (cylindrical due to the azimuthal symmetry of the problem) equation, describing drift and diffusion of excitons in TMDCs \cite{Harats2020DynamicsWS2b}:
\begin{multline}
\label{eq:diff_drift}
\nabla(D\nabla n(r))+\nabla(\mu\; n(r)\nabla u(r))- \\
\frac{n(r)}{\tau}-n^2(r)R_A+S(r)=0
\end{multline}
where $D$ is the diffusion coefficient, $\mu$ is the exciton mobility related to the diffusion coefficient through the Einstein relations ($\frac{D}{\mu}=k_BT$ where $T$ is the temperature), $\tau$ is the exciton lifetime, $R_A=0.14\;cm^2/s$ is the Auger recombination rate \cite{Kulig2018}, and $S(r)$ is the excitation profile.

We consider two different excitation profiles $S(r)$. The first profile describes a laser beam focused into a diffraction-limited spot and is similar to the one used in our previous work \cite{Harats2020DynamicsWS2b}:
\begin{equation}\label{eq:laser_exc}
    S_{laser}(r)=\frac{\alpha}{\hbar \omega}\frac{I_0}{2\pi\sigma^2}e^{-\frac{r^2}{2\sigma^2}}
\end{equation}
 where $I_0=8\;nW$  and $\sigma=600nm/2\sqrt{2\ln 2}$ are the parameters characterizing the power and the width of the laser spot in our previous experiment  \cite{Harats2020DynamicsWS2b}, while $\alpha =5\;\%$ and $\hbar \omega$ are the absorption coefficient of a TMDC and the photon energy, respectively. This profile is useful in generating large densities of excitons in the central area and in studying excitonic transport.
 
 The second excitation profile mimics uniform solar illumination:
\begin{equation}\label{eq:solar_exc}
    S_{solar}(r)=\left\{
        \begin{array}{c c}
           \frac{\alpha}{\hbar \omega} I_{solar} & r \leq r_{mem}\\
        0 & r_{mem} \leq r
        \end{array}\right.
\end{equation}
where $I_{solar}=1000\;W/m^2$ is the solar intensity. Note that the solar flux is 2 order of magnitude lower than the laser intensity, due to the tight laser spot compared to the membrane size. This profile is used to compare our results with the original proposal of Feng et al. in which similar devices were suggested for photoconversion \cite{Feng2012}. For this profile, we also assume a spectrally uniform light absorption of $\alpha=5\;\%$. We note that while this assumption is obviously a gross over-simplification for the case of the solar spectrum, our results depend on that value only weakly. We also neglect rather weak changes of the absorption profile with strain.

Finally, we use all of the assumptions above to find the spatially-dependent exciton density $n(r)$ by solving Eq. \ref{eq:diff_drift}. We investigate  the dependence of $n(r)$ on several physical parameters - the temperature $T$, the diffusion coefficient $D$ (and the mobility $\mu$ related to $D$ through the Einstein relations), the lifetime $\tau$, and the excitation profile $S(r)$. 


\section{Results}
 We define the efficiency of a funneling device as the fraction of all photo-generated excitons reaching the central tip:
\begin{equation}\label{eq:efficiency}
    eff=\frac{\int_0^{r_{tip}}n(r)rdr}{{\int_0^{r_{mem}}n(r)rdr}}
\end{equation}

We start by examining the funneling efficiency at room temperature $T=300\;K$. The primary motivation for this calculation comes from the unexpected experimental observation of low funneling efficiency ($\sim 4\;\%$) under these conditions \cite{Harats2020DynamicsWS2b}. We take the exciton lifetime to be $\tau=1\;ns$, around the measured value for a typical TMDC (WS$_2$) at room temperature \cite{Kulig2018}. The diffusion coefficient $D$ characterizes the sample quality and depends on the defect density. Experimentally, the diffusion coefficient was found to be $D=0.3\;cm^2/s$ for monolayer WS$_2$ \cite{Kulig2018} at room temperature. For hBN-encapsulated TMDCs, a significantly higher value, $D=10\;cm^2/s$ has been reported \cite{Zipfel2020ExcitonDisorder}. 
To account for variations in possible disorder levels, we plot the efficiency for values of $D$ spanning five orders of magnitude. 

\begin{figure}[hbtp]
    \centering
    \includegraphics[width=0.9\columnwidth]{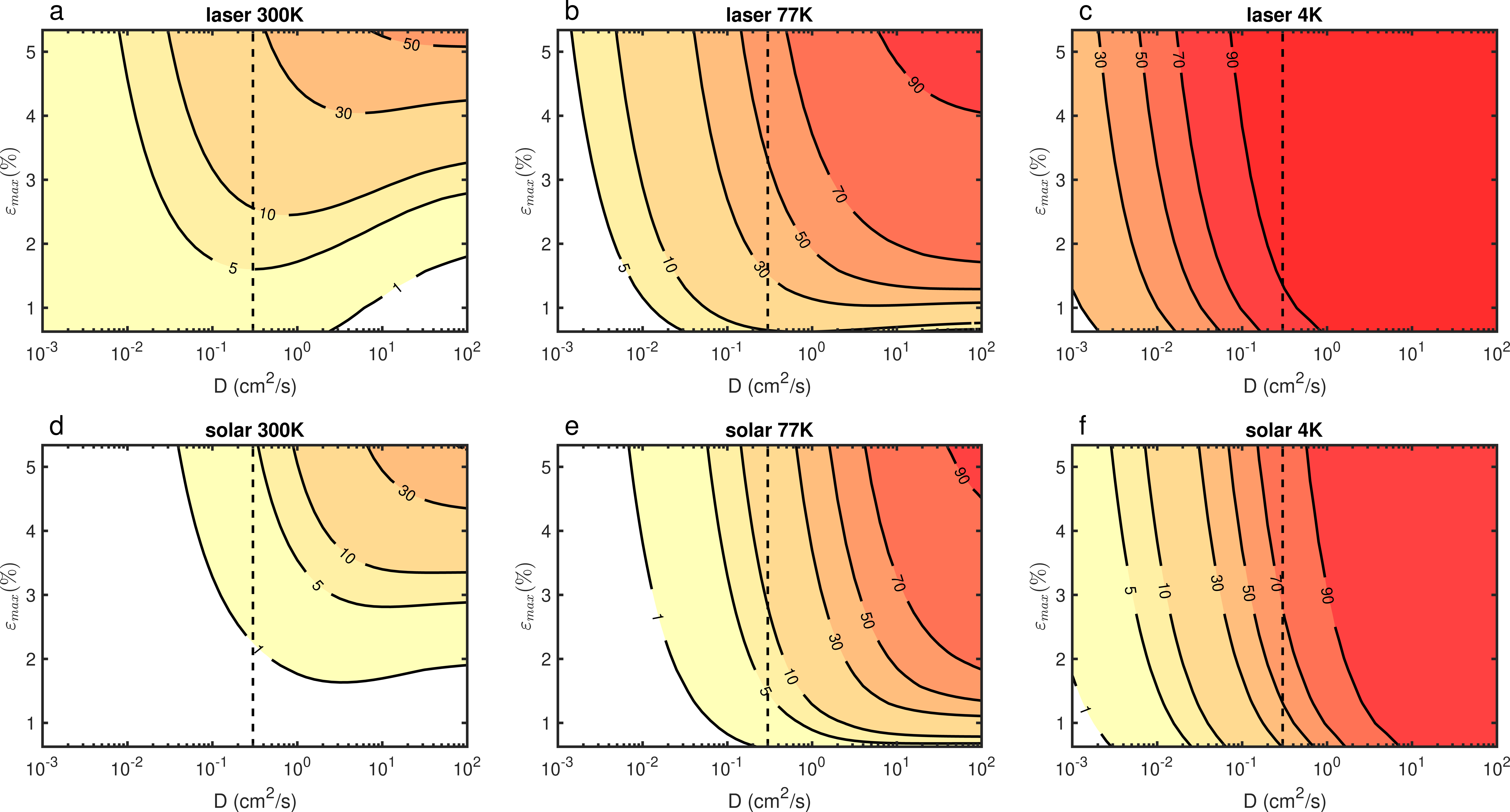}
    \caption{Contour plots of the funneling efficiency as a function of the diffusion coefficient $D$ and the maximum strain value $\varepsilon_{max}$ at various temperatures. The panels (a)-(c) correspond to the laser illumination profile (Eq. \ref{eq:laser_exc}), while the panels (d)-(f) correspond to the solar illumination profile (Eq. \ref{eq:solar_exc}). The vertical black dashed line corresponds to the experimental room temperature diffusion coefficients $D=0.3\;cm^2/s$. The exciton lifetime is assumed to be constant $\tau=1\; ns$ in this plot.}
    \label{fig:fig2_temp_change}
\end{figure}

Figures \ref{fig:fig2_temp_change}a,d show the funneling efficiency given by Eq. \ref{eq:efficiency} at room temperature (colorscale) for both solar and laser illumination profiles (Eqs. \ref{eq:laser_exc}, \ref{eq:solar_exc}). It is plotted as a function of the height of the central pillar, parameterized by the maximum strain $\varepsilon_{max}$, and the disorder level parameterized by $D$. For small $\varepsilon_{max}$ or small $D$ the funneling efficiency is close to zero. In that regime, excitons are essentially immobile: their motion due to both diffusion ($l_{diff}=\sqrt {D\tau}$) and drift ($l_{drift} \propto \mu \tau \nabla u$) during their lifetime is negligible compared to the membrane size $r_{mem}$. The fraction of excitons reaching the tip is then the same as the fraction of excitons created in the tip area to the whole excitation area. Since the solar excitation profile is spatially broader in comparison with the laser excitation profile, funneling in this profile is less efficient.

The drift length $l_\mu$ is enhanced as we increase either the strain (and $\nabla u$) or the diffusion coefficent $D$. Hence, the funneling efficiency also grows with both strain and $D$ (Figs. \ref{fig:fig2_temp_change}a,d). However, for the diffusion coefficient value corresponding to previously measured monolayer WS$_2$ (vertical dashed line in \ref{fig:fig2_temp_change}), the efficiency remains low, less than 30\%, for both excitation profiles at room temperature. This is consistent with our previous experimental observations \cite{Harats2020DynamicsWS2b}. There are two complementary reasons for that. First, the excitons do not live long enough to be transported towards the center of the membrane. Second, the diffusion dominates over the drift in these devices \cite{Harats2020DynamicsWS2b}. 

Next, we consider the influence of the temperature on the funneling efficiency. The primary motivation for this, is that at low temperatures we expect the role of the drift term to grow in Eq. \ref{eq:diff_drift} compared to the diffusion term and we therefore expect higher funneling efficiency. The analysis of that equation shows two distinct mechanisms through which the temperature can affect funneling efficiency. First, the temperature determines the ratio between the exciton mobility $\mu$ (the drift term) and the diffusion coefficient $D$ due to the Einstein relations. Second, the material parameters such as the diffusion coefficient $D$ and the exciton lifetime $\tau$ are themselves temperature-dependent.  For excitons in TMDC, the lifetime decreases  from $ns$ to $ps$ between room and cryogenic temperatures \cite{Robert2016ExcitonMonolayers,Palummo2015ExcitonDichalcogenides}. The temperature dependency of the diffusion coefficient is still unknown. Intuitively, the defect-related localization of excitons should lead to the decrease of $D$ at cryogenic temperatures. Nevertheless, a recent theoretical prediction suggests that the diffusion coefficient may in fact increase due to quantum effects \cite{Glazov2020QuantumSemiconductors}. To generalize our discussion, we first focus on the temperature-dependence of the funneling efficiency stemming from the drift/diffusion ratio changes while fixing the exciton lifetime to $\tau=1\;ns$. We will then consider separately the effect of temperature-dependence on the lifetime.

Figure \ref{fig:fig2_temp_change} shows the funneling efficiency for both excitation profiles and different temperatures. The temperatures were chosen to correspond to common experimental conditions - $4\;K$, $77\;K$ and $300\;K$ for experiments in liquid $He$, liquid $N_2$, and room temperature, respectively. The contour plots in Fig. \ref{fig:fig2_temp_change} show that as the temperature is decreased, the efficiency strongly increases for both excitation profiles. This is not surprising, as the the drift term begins to dominate over the diffusion, in agreement with the Einstein relations. At 4K, we observe  efficient ($>50\;\%$) funneling for $D=0.3\;cm^2/s$ for relatively low strains $\varepsilon_{max}\sim 1.5\;\%$. Nevertheless, for very low diffusion coefficients, $D\sim 10^{-3}\;cm^2/s$, the efficiency remains  low even for high maximum strain $\varepsilon_{max}$. In this regime, excitons "freeze-out" so that the funneling efficiency in negligible.

\onecolumngrid

\twocolumngrid

\begin{figure}[hbtp]
    \centering
    \includegraphics[width=0.9\columnwidth]{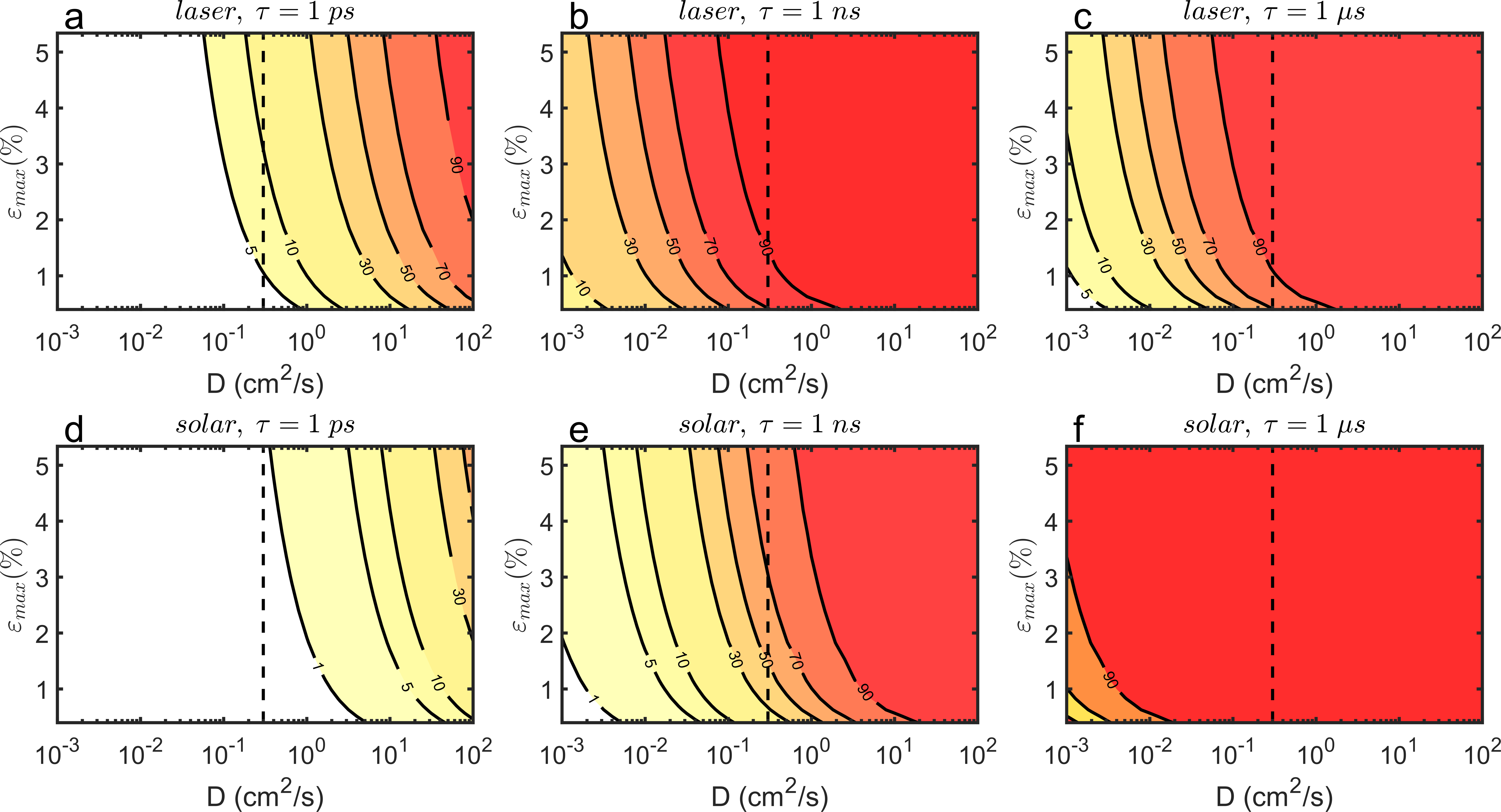}
    \caption{Funneling efficiency at cryogenic temperature $T=4\;K$. Funneling efficiency is shown as contour plots as a function of the diffusion coefficient $D$ and the maximum strain value $\varepsilon_{max}$ for various exciton lifetimes. The panels (a)-(c) correspond to the laser illumination profile (Eq. \ref{eq:laser_exc}), while the panels (d)-(f) correspond to the solar illumination profile (Eq. \ref{eq:solar_exc}). For all graphs the vertical black dashed line corresponds to the reported diffusion coefficient value at room temperature $D=0.3\;cm^2/s$ \cite{Kulig2018}.}
    \label{fig:fig3_LT}
\end{figure}

Next, we consider the variation of the funneling efficiency with the exciton lifetime. We concentrate on the case of cryogenic temperature $T=4\;K$ which corresponds to the maximized efficiency (Fig. \ref{fig:fig2_temp_change}c,f). We consider the lifetimes between $\tau=1\;ps$, closer to the near-radiative lifetime experimentally reported for monolayer WS$_2$ at 4K for excitons \cite{Robert2016ExcitonMonolayers,Palummo2015ExcitonDichalcogenides} and $\tau=1\;\mu s$, on the order of the lifetime of indirect excitons in hetero-bilayers such as WSe$_2$/MoSe$_2$ \cite{Nagler2017InterlayerHeterostructure,Miller2017Long-LivedHeterostructures,Jiang2018MicrosecondHeterostructures}. A very strong increase of the funneling efficiency with increasing $\tau$ is obvious in Fig. \ref{fig:fig3_LT}. While for $\tau=1\;ps$ the efficiency is small throughout the parameter space (Fig. \ref{fig:fig3_LT}a,d), values close to $100 \;\%$ are achieved in the same region of parameters for $\tau=1\;\mu$s (Fig. \ref{fig:fig3_LT}c,f). Nearly all excitons reach the funnel center in this regime. This behaviour is expected: while the diffusion length has a square-root dependency on the lifetime $\tau$ ($l_D=\sqrt{D\tau}$), the drift length depends on it linearly ($l_\mu\propto \mu\tau$ \cite{Harats2020DynamicsWS2b}). Therefore, the relative role of drift and hence funneling efficiency increases for longer $\tau$.

Closer examination of Fig. \ref{fig:fig3_LT} reveals one surprising feature. The funneling efficiency is much higher for the case of the solar (Fig. \ref{fig:fig3_LT}f) compared to the laser excitation profile (Fig. \ref{fig:fig3_LT}c) when $\tau=1\;\mu s$ and all other parameters are the same. This is exactly the opposite compared to the behaviour seen at room temperature for $\tau=1\;ns$ (see Fig. \ref{fig:fig2_temp_change}c,f). This behaviour stems from the Auger recombination term $R_A$ in Eq. \ref{eq:diff_drift}. During the funneling process, a large density of excitons generated throughout the device is transported to a relatively small region defined by the tip. Alas, as the excitons concentrate in the funnel, they recombine non-radiatavely via Auger recombination \cite{Kulig2018}. This results in the reduction of the funneling efficiency. As the laser excitation intensity is at least 2 orders of magnitude higher than the solar excitation intensity, the relative Auger recombination rate following Eq. \ref{eq:auger_cond} -- and hence the reduction of the funneling efficiency -- is more significant for the laser excitation profile.

We can estimate the importance of Auger recombination in our devices. For this recombination to occur, a pair of excitons must be in close proximity during its' lifetime. According to Eq. \ref{eq:diff_drift}, the Auger term is negligible compared to the radiative recombination term when 
\begin{equation}\label{eq:auger_cond}
    n(0)\tau R_A \ll 1
\end{equation}
where $n(0)$ is the exciton density at the funnel center. As long as this condition is satisfied, the efficiency is maximized. For the case of the solar profile, Eq. \ref{eq:auger_cond} is always satisfied (even for $\tau=1\;\mu s$) whereas for the laser profile it is satisfied only for very short lifetimes ($\tau=1\;ps$). We therefore conclude that the Auger recombination limits the efficiency for the high intensity laser illumination. This limitation is unimportant for the solar excitation profile.

\begin{figure}[hbtp]
    \centering
    \includegraphics[width=0.9\columnwidth]{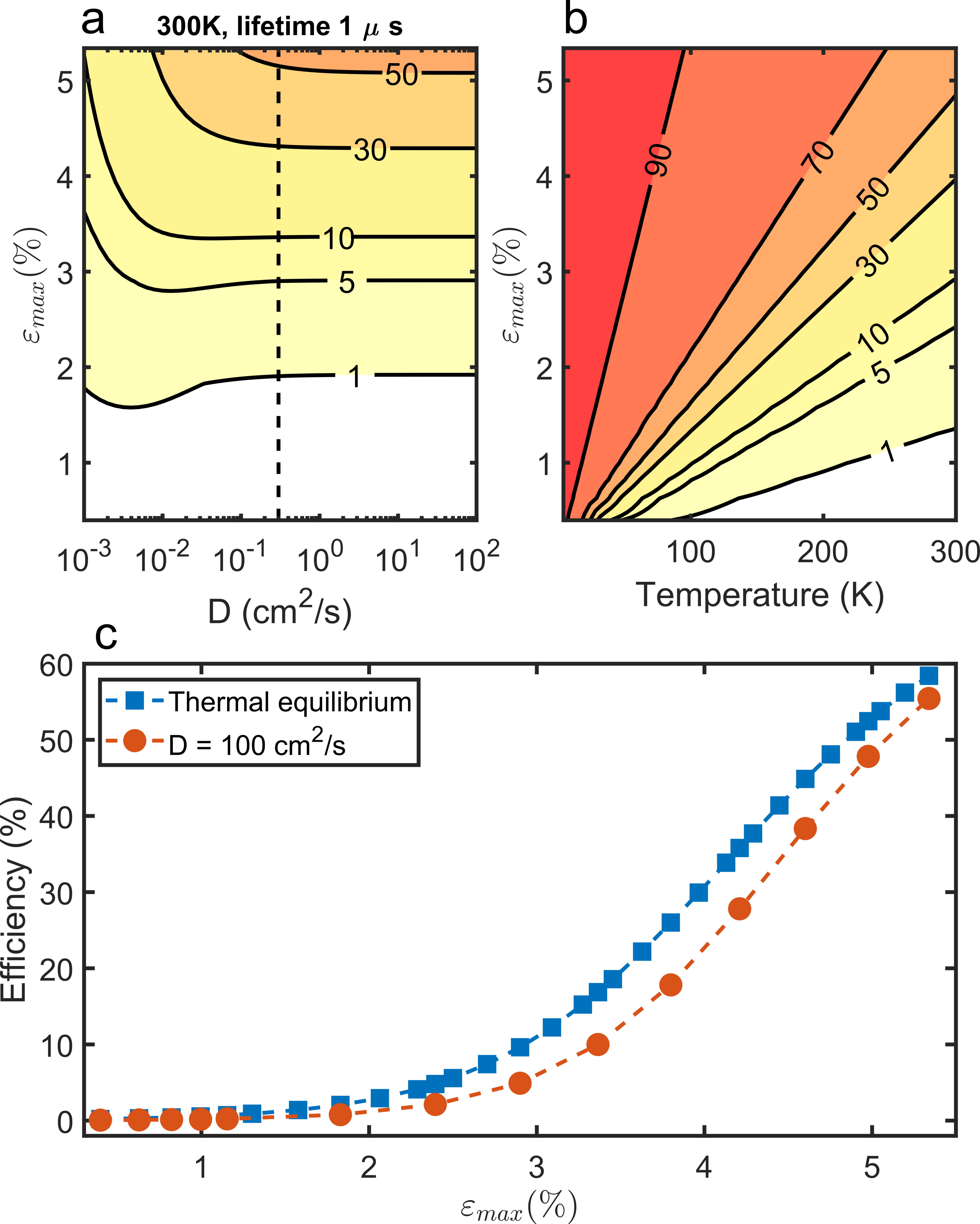}
    \caption{(a) The calculated efficiency for solar excitation profile at room temperature with $\tau=1\;\mu s$. (b) The maximum efficiency as a function of the maximum strain and the temperature, derived from the thermal equilibrium condition (Eq. \ref{eq:max_eff}). (c) Comparison of the efficiency calculated at thermal equilibrium for $T=300\;K$ from (b) (blue squares) and at $D=10^2\;cm^2/s$ from (a) (red circles). }
    \label{fig:fig4_thermal_eq}
\end{figure}

So far, we found out that the funneling can be a very efficient process at low temperature as long as the excitonic lifetime is long enough and the illumination intensity is small enough for Auger recombination to be negligible. How would a device with the same parameters function at room temperature? In Fig. \ref{fig:fig4_thermal_eq}a, we show the funneling efficiency for a lifetime of $\tau=1\;\mu s$ at $T=300\;K$ for the solar excitation profile. While, as expected, funneling is less efficient compared to the case of cryogenic temperatures (Fig. \ref{fig:fig3_LT}f), the efficiency is significantly higher than the experimental value of $\sim 4\;\%$ \cite{Harats2020DynamicsWS2b}. It reaches more than $50 \;\%$ at $\varepsilon_{max} \sim 5 \;\%$ and $D=0.3\;cm^2/s$. It is interesting to note that while for the previously considered case of short lifetimes we observe that the efficiency grows with $D$, for the long-lifetime device shown in Fig. \ref{fig:fig4_thermal_eq}a, it is nearly $D$-independent for values of $D>1\;cm^2/s$. This is the sign that the excitons now live long enough to reach thermal equilibrium in the funnel. Indeed, in thermal equilibrium the probability of the excitons to occupy a certain portion of the funnel is only dependent on the energy of that state and not on the material parameters such as the diffusion coefficient or the lifetime.

\onecolumngrid

\twocolumngrid

To check this hypothesis, we calculated the funneling efficiency of the device at thermal equilibrium. In equilibrium, the probability of an exciton to reside at a specific position is given by the Boltzmann distribution. The efficiency at thermal equilibrium is as follows:
\begin{equation}\label{eq:max_eff}
    eff^{max}=\frac{\int_0^{r_{tip}} e^{-\frac{\Delta u(r)}{k_BT}}rdr}{\int_0^{r_{mem}} e^{-\frac{\Delta u(r)}{k_BT}}r dr}
\end{equation}
where $\Delta u(r)=E_0-u(r)$ is the difference between the local  band-gap energy and its unstrained value.

Figure \ref{fig:fig4_thermal_eq}b shows the  efficiency calculated from Eq. \ref{eq:max_eff} as a function of strain and temperature. We then compare the efficiency extracted from that graph at room temperature (Fig. \ref{fig:fig4_thermal_eq}c, blue symbols) to the efficiency calculated previously for the device with long lifetime $\tau=1\;\mu$s and high diffusion coefficient $D=10\;cm^2/s$ under solar illumination profile (Fig. \ref{fig:fig4_thermal_eq}c, red symbols). We see that the curves are similar. This confirms that for long exciton lifetime $\tau=1\;\mu s$, thermal equilibrium is nearly reached.  Moreover, Eq. \ref{eq:max_eff} presents a simple approach to evaluate the maximum possible efficiency of the funneling device regardless of material parameters. The limit is approached when the lifetime is long enough and Auger recombination rate is small enough. We confirm that at a realistic strain of $5 \;\%$ the limit of the efficiency at room temperature is about $50 \;\%$. At cryogenic temperatures, $T<77$ K, is is possible to achieve $>90 \;\%$ even at relatively small strain values (Fig. \ref{fig:fig4_thermal_eq}b).

\section{Conclusions}
We investigated the efficiency of the exciton funneling process in non-uniformly strained TMDC. The fundamental limitation for the funneling efficiency is set by the balance between drift and diffusion which is achieved in thermal equilibrium. At room temperature this limit is about $50 \;\%$ at a maximum strain of $5\;\%$. This limit can be approached in devices with sufficiency high diffusion coefficients, long enough exciton lifetimes and under weak illumination, avoiding Auger recombination. While typical TMDC monolayers do not meet these requirements at room temperature, the limit is nearly reached in hetero-bilayer TMDC device with long-lived ($\approx 100\;ns$) interlayer excitons. At cryogenic temperatures, the excitonic drift is favored over diffusion which leads to much more efficient funneling for the same material parameters, reaching near-unity values. We should stress that the temperature dependence of the material parameters such as $D, R_A$, and $\tau$ is not fully known yet, making it hard to make specific prediction for that regime. At the same time, we expect that the Auger recombination rate will decrease as a function of temperature as it is a phonon-assisted process \cite{Guichard2008Temperature-dependentNanowires,Kharchenko1996AugerWells}. Similarly, there is a prediction for higher diffusion coefficient at low temperature \cite{Glazov2020QuantumSemiconductors} which should increase the efficiency. We therefore expect that our estimates for the low-temperature regime are robust as well. The devices considered here are interesting for efficient collection of photogenerated excitons. While the original proposal considered the application in photoconversion, it may also be especially interesting to consider the use of this approach to generate large excitonic densities in the central region of the funnel. A plethora of correlated phenomena including the Bose-Einstein condensation and superfluidity is expected and has been partially observed in that regime at low temperatures \cite{Wang2019EvidenceLayers,Sigl2020CondensationHeterostack}, making the investigation of that regime worthwhile.

\section{Acknowledgments}

We acknowledge enlightening conversations with Roland Netz and Sebastian Heeg. This work was supported by the European Research Council Starting grant 639739, and DFG CRC/TRR 227.

\bibliographystyle{apsrev4-2.bst}
\bibliography{references}

\end{document}